%% file: schw_simulation_letter.tex
\documentclass[a4paper,twoside]{article}

\input{macrosquist}\preprintsize

\usepackage{amsmath}
\usepackage{amssymb}
\usepackage{cmmib57}
\usepackage{graphicx}

\acronym{\SSS}{SSS}{\emph{static spherically symmetric}}
\acronym{\MF}{MF}{Matter Frame}
\acronym{\LF}{LF}{Lab Frame}

\begin{document}
\bibliographystyle{prsty}

\reportnumber{USITP-03-09\\[2pt]\eprintstyle{gr-qc/0309104}\\[2pt]22-sep-2003}

\title{A moving medium simulation of Schwarzschild black hole optics}
\author{Kjell Rosquist\footnote{\email{kr@physto.se}} \\
        \small{Department of Physics, AlbaNova University Center,} \\[-7pt]
        \small{Stockholm University, 106 91 Stockholm, Sweden}}
\date{}

\maketitle
        
\begin{abstract}
\begin{center}
\begin{minipage}[t]{0.8\linewidth}\small{
        An explicit fluid flow simulation of electromagnetic wave propagation
        in the gravitational field of a Schwarzschild black hole is given. The
        fluid has a
        constant refractive index and a spherically symmetric inward
        directed flow. The resulting form of the metric leads to a new
        coordinate system in which the Schwarzschild vacuum is written in
        Gordon's form. It is shown that a closely related coordinate system
        interpolates between the Kerr-Schild and Painlev\'e-Gullstrand
        coordinates.}
\end{minipage}
\end{center}
\end{abstract}

\vspace{1cm}

\centerline{\bigskip\noindent PACS numbers: 04.70.Bw, 04.70.Dy, 04.80.-y}
\centerline{\bigskip\noindent Keywords: Analogue model; black hole; moving medium}

\section{Introduction}

Analogue models of gravitational fields have been rather extensively 
discussed in recent years (see \emph{e.g.\/} \cite{Novello_etal:2002}). One motivation is that it might become possible in the near future to simulate kinematic aspects of strong gravitational fields. A number of different ways of achieving this goal have been proposed, mainly by using acoustic or electromagnetic waves. An effect which could in principle be measured by simulating a black hole would be the Hawking radiation associated with the black hole horizon \cite{Visser:2003}. An encouraging development is the recent experiments on media with a very large effective index of refraction \cite{Hau_etal:1999}. Since the light speeds in these experiments are of the order of meters or centimeters per second, it would be possible to simulate an horizon which could lead to observable Hawking radiation. The horizon will be the boundary of the region where the fluid velocity is larger than the velocity of light in the medium.
A complication is that the low values of the light velocity are due to a large dispersion. Therefore, it will not be possible in general to give a geometric formulation of the dynamics of light propagation in such a situation. However, if the dispersion is sufficiently ``well-behaved", it seems that one can still define an effective metric for light propagation \cite{Leonhardt&Piwnicki:2000}. In this note we will consider only the more simple dispersion-free case.

We will present an explicit theoretical model for simulating the Schwarzschild vacuum field, using a (hypothetical) dispersion-free transparent dielectric fluid. The refractive index is assumed to be independent of the density. The fluid flow is spherically symmetric and inward directed. A byproduct of this work is a new representation of the Schwarzschild gravitational field in terms of the Gordon form corresponding to this flow. The metric form given by Gordon \cite{Gordon:1923} is the natural representation of the (simulated) gravitational field in this context. It is this particular way of writing the metric which results in a new coordinate system for the Schwarzschild vacuum. Like some other coordinate systems (\emph{e.g.\/} Kruskal \cite{Misner_etal:1973}, Painlev\'e-Gullstrand \cite{Robertson&Noonan:1968,Martel&Poisson:2001} and Kerr-Schild \cite{Misner_etal:1973}) it is non-singular at the horizon and can be continued all the way down to the central singularity. Another nice property of this representation of the gravitational field is that it can be viewed as a finite perturbation of the Minkowski space-time. It shares this property with the Kerr-Schild form of the metric. However, it has the additional advantage of having the direct physical interpretation in terms of the flowing dielectric fluid.

Consider now a transparent fluid moving with 4-velocity $u^\mu$ in our laboratory. In practical applications, the velocity of the fluid will always be non-relativistic. It is nevertheless useful to use a relativistic notation in this context. The lab geometry is given by the usual Minkowski metric, here shown in Lorentzian coordinates $x^\mu = (ct, x, y , z)$
\begin{equation}
   \eta = -c^2 \d t^2 + \d x^2 + \d y^2 + \d z^2 \ .
\end{equation}
If the fluid has a refractive index, $n$, which is constant or slowly varying (in space and time), then the propagation of electromagnetic waves is governed by a wave operator
\begin{equation}
   g^{\mu\nu} \frac{\partial^2}{\partial x^\mu \partial x^\nu} \ ,
\end{equation}
where $g^{\mu\nu}$ is Gordon's metric \cite{Gordon:1923}
\begin{equation}
   g^{\mu\nu} = \eta^{\mu\nu} - m\, u^\mu u^\nu \ ,
\end{equation}
with $m=n^2-1$. The covariant form of Gordon's metric is
\begin{equation}
   g_{\mu\nu} = \eta_{\mu\nu} + k\, u_\mu u_\nu \ ,
\end{equation}
where $k=m/n^2 = 1-n^{-2}$ and\footnote{Since we are dealing with two different geometries, $\eta$ and $g$, it is important to specify which metric is involved in raising and lowering of indices.}  $u_\mu = \eta_{\mu\nu} u^\nu$.

\section{Simulating the Schwarzschild vacuum geometry by a moving fluid}

We will now show how the Schwarzschild vacuum can be simulated by a spherically symmetric fluid flow. The Minkowski geometry in spherical coordinates with origin at the center of the fluid flow has the form
\begin{equation}\label{eq:labmetric}
  \eta = -c^2 \d t^2 + \d r^2 + r^2 \d\Omega^2 \ .
\end{equation}
where as usual
\begin{equation}
   \d\Omega^2 = \d\theta^2 + \sin^2\!\theta\,\d\phi^2
\end{equation}
is the 2-sphere metric. The fluid 4-vector can then be written in the form $u^\mu= (\cosh\zeta, -\sinh\zeta, 0, 0)$ where $\zeta= \arctanh(v/c) = \zeta(r)$ is the boost parameter and $v= |\bfv| = -v^r$ is the velocity.  We have chosen a minus sign for the radial component of $u^\mu$ to make the fluid flow inward directed. The Gordon metric becomes
\begin{equation}
    g = \eta +k(\cosh\zeta\, c\,\d t + \sinh\zeta\, \d r)^2 \ .
\end{equation}
It turns out that this metric will represent a Schwarzschild black hole with mass $M$ if the fluid has a constant refractive index and its velocity is given by
\begin{equation}\label{eq:v}
    v= \frac{c}{\sqrt{1+(n^2-1)\cdot\displaystyle\frac{r}{r_\text{g}}}} \ ,
\end{equation}
where $r_\text{g} = 2GM/c^2$ is the gravitational radius.
The corresponding fluid 4-velocity has the components
\begin{equation}
    u^t = \sqrt{\frac{n^2-f}{n^2-1}} \ ,\qquad
    u^r = -\sqrt{\frac{1-f}{n^2-1}} \ ,
\end{equation}
where $f=1-r_\text{g}/r$. The velocity of light in the medium is $c_\text{m} = c/n$. Comparing this with the fluid velocity itself, we find as expected that the fluid becomes superluminal at $r=r_\text{g}$, \emph{i.e.\/} at the black hole horizon. Of course the problem of the physical realization of the associated fluid flow remains to be addressed.  The only possibility for an experimental realization of the horizon seems to be to have $n$ large.  In that case we can make the approximation $v\approx c_{\text{m}} \sqrt{r_\text{g}/r}$.  As it turns out, if the speed of sound of the fluid is constant, this flow also generates an acoustic black hole with a conformally Schwarzschild metric \cite{Visser:1998}. To see what the flow given by equation \eqref{eq:v} is like we must solve the continuity equation for a stationary spherically symmetric fluid
\begin{equation}
    \nabla\cdot (\rho \bfv) = \frac{\d}{\d r}(r^2 \rho v) = 0 \ .
\end{equation}
This gives the radial dependence of the density as
\begin{equation}
    \rho = \frac{K}{r^2}\sqrt{1+(n^2-1)
          \cdot\displaystyle\frac{r}{r_\text{g}}} \ ,
\end{equation}
where $K$ is an integration constant. We see that the flow is compressible.

We have defined a fluid flow which gives rise to an effective dielectric metric which coincides with the Schwarzschild vacuum geometry. To make this more clear we write down the solution in its explicit form (from now on setting $c=G=1$)
\begin{equation}\label{eq:metric}
    g = -n^{-2}f\,\d t^2 + 2n^{-2}\sqrt{(1-f)(n^2-f)}\,\d t\,\d r
             +[1+n^{-2}(1-f)]\,\d r^2 + r^2 \d\Omega^2 \ ,
\end{equation}
recalling that $f=1-2M/r$ and that $n$ is constant. The usual form of the Schwarzschild metric can be recovered by the coordinate transformation
\begin{equation}\label{eq:dt}
  \d t = n\,\d t_S+f^{-1}\sqrt{(1-f)(n^2-f)}\,\d r \ ,
\end{equation}
where $t_{\text{S}}$ is the usual Schwarzschild time coordinate. The metric then takes the familiar form
\begin{equation}
   g = -f\,\d t_\text{S}^2 + f^{-1} \d r^2 + r^2 \d\Omega^2 \ .
\end{equation}
For reference we also give the integrated form of the transformation \eqref{eq:dt}
\begin{equation}
    t = n\,t_{\text{S}} + \frac{4M\sqrt{n^2-f}}{\sqrt{1-f}}
       + 2Mn\ln\left\{\frac
        {\bigl(1-\sqrt{1-f}\bigr)\bigl(n^2+n\sqrt{n^2-f}\bigr)-f}
        {\bigl(1+\sqrt{1-f}\bigr)\bigl(n^2+n\sqrt{n^2-f}\bigr)-f} \right\} \ .
\end{equation}
At infinity, where $f=1$, the metric \eqref{eq:metric} reduces to
\begin{equation}\label{eq:ginfinity}
    g = -n^{-2}\d t^2 + \d r^2 + r^2 \d\Omega^2 \ .
\end{equation}
This is the flat Minkowski space-time as it should be. The time component of the metric is rescaled accounting for the slower speed of light in the medium.

We conclude this section by showing that these coordinates are related to both the Kerr-Schild and the Painlev\'e-Gullstrand coordinates. First, we note that the metric \eqref{eq:metric} is really a family of metrics parametrized by $n$. As it appears in \eqref{eq:metric}, the metric is defined for all values of $n$ in the range $n\geq1$, even though the underlying simulation is only defined for $n>1$. Setting $n=1$, the metric becomes
\begin{equation}
   g = \eta + \frac{2M}{r}(\d t + \d r)^2 \ ,
\end{equation}
which is the Kerr-Schild form of the Schwarzschild geometry. This is very natural since the fluid velocity tends to the speed of light in the limit $n \rightarrow 1$ as can be seen from \eqref{eq:v}. The metric also has a limit as $n \rightarrow \infty$. To perform this limit we must first introduce a new time variable $T=t/n$ in \eqref{eq:metric}. The metric then reads
\begin{equation}
   g = -\left(1-\frac{2M}{r}\right)\d T^2
      + 2\sqrt{\frac{2M}{r}\left(1-\frac1{n^2}+\frac{2M}{n^2r}\right)}\d T\d r
      + \left(1+\frac{2M}{n^2r}\right)\d r^2
      + r^2 \d\Omega^2 \ .
\end{equation}
It is now possible to take the limit $n \rightarrow \infty$ and the result is
\begin{equation}
   g = -\left(1-\frac{2M}{r}\right)\d T^2 + 2\sqrt{\frac{2M}{r}}\,\d T\d r
      + \d r^2 + r^2 \d\Omega^2 \ .
\end{equation}
This is just Schwarzschild's metric in Painlev\'e-Gullstrand coordinates. Finally, introducing the parameter $s=1-k=n^{-2}$ taking values in the interval $0\leq s \leq1$, the metric becomes
\begin{equation}
   g = -f\, \d T^2 + 2\sqrt{(1-f)(1-sf)}\d T\d r
      + (1+s-sf)\d r^2 + r^2 \d\Omega^2 \ .
\end{equation}
This metric form interpolates between the Painlev\'e-Gullstrand coordinates at $s=0$ and the Kerr-Schild coordinates at $s=1$.

\section{Discussion}
In the metric given by \eqref{eq:metric}, the interpretation of the time coordinate $t$ is that it is the time of the lab in which the simulation is taking place. This time is different from the Schwarzschild time $t_\text{S}$. The radial variable $r$ plays two roles. First, it is the Euclidean radial distance in the lab. But it is also the Schwarzschild radial variable in the simulated black hole. Another thing to note about the simulation is that there are actually two background geometries, one which is the background in the absence of the medium with metric \eqref{eq:labmetric}. This is the lab geometry. There is also the background in the medium at rest with metric \eqref{eq:ginfinity}. Both these backgrounds are flat and they are related by the time transformation $t = nT$. We can interpret $T$ as an ``electromagnetic time" in the medium at rest. This means that a clock based on light travel time which shows the time $t$ outside the medium would show the ``slower" time $T$ inside the medium.

As discussed in the introduction, the Gordon form of the metric is interesting in its own right. Having found the flow corresponding to a Schwarzschild black hole, a natural question to ask, is if it would be possible to do the same thing with Kerr's rotating black hole. In a wider perspective one could ask which space-times admit a Gordon representation. The model presented in this letter has the property that the refractive index of the fluid is constant. However, models with a non-constant $n$ are also possible. A physical requirement for such models to serve as simulations would be the existence of a constitutive relation $n=n(\rho)$. In the spherically symmetric case this entails no restriction since there is only one non-trivial variable, and a constitutive relation is then always given parametrically by $n=n(r)$ and $\rho=\rho(r)$. In situations with more than one non-trivial variable, however, one would need to consider the Einstein equations coupled to the continuity equation.

A more complete account of this work, including the method of finding the solution, will be given in a separate report.

\bibliography{kr}
\end{document}

%% file: macrosquist.tex
\newcount\commentcount \commentcount=0 
 \long\def\comment#1{\ifnum\commentcount=1 #1\fi}

\newcommand{\preprintsize}{
      \headheight=0pt                               
	  \topmargin=0.5cm \headsep=1.5cm               
      \oddsidemargin=-0.5cm \evensidemargin=-0.5cm  
      \textheight=22truecm \textwidth=17truecm      %
	  \setlength{\columnsep}{20pt}                  
}

\newtoks\reportnoregister \newtoks\eprintnoregister
\newcommand{\reportnumber}[1]{\reportnoregister={#1}}
\newcommand{\eprintnumber}[1]{\eprintnoregister={#1}}

\reportnumber{\mbox{}} 
\eprintnumber{\mbox{}} 

\newcommand{\reportid}{
   \begin{minipage}{17cm}\vspace{-2.6cm}  
     \begin{flushright}
      {\normalsize \the\reportnoregister \\[-.2cm]
	    \eprintstyle{\the\eprintnoregister}}\vspace{2.6cm} 
     \end{flushright}
   \end{minipage}\hspace{-17cm} }

\catcode`@=11   
\def\title#1{\gdef\@title{\reportid#1}}
\catcode`@=12   

\newcommand{\email}[1]{E-mail: \eprintstyle{#1}}
                                   
\newcommand{\eprintstyle}[1]{\textsf{#1}} 

\newcommand{\journalfont}{\rm}  
\newcommand{\jou}[1]{{\journalfont #1\ }}
\newcommand{\joudef}[2]{\newcommand #1{\jou{\ignorespaces #2}}}

\joudef{\ajp}    { Am.~J.~Phys.}
\joudef{\aaa}    { Astron.\ Astrophys.}
\joudef{\aip}    { Adv.\ Phys.}
\joudef{\adm}    { Adv.\ Math.}
\joudef{\am}     { Ann.\ Math.}
\joudef{\apb}    { Ann.\ Phys.\ (Berlin)}
\joudef{\apny}   { Ann.\ Phys.\ (N.Y.)}
\joudef{\apj}    { Astrophys.\ J.}
\joudef{\apjs}   { Astrophys.\ J.\ Suppl.}
\joudef{\cjp}    { Can.\ J.\ Phys.}
\joudef{\cmp}    { Commun.\ Math.\ Phys.}
\joudef{\cqg}    { Class.\ Quantum Grav.}
\joudef{\faa}    { Funct.\ Anal.\ Appl.}
\joudef{\grg}    { Gen.\ Rel.\ Grav.}
\joudef{\ijmpd}  { Int.\ J.\ Mod.\ Phys.\ D}
\joudef{\ijtp}   { Int.\ J.\ Theor.\ Phys.}
\joudef{\invm}   { Invent.\ Math.}
\joudef{\jm}     { J.\ Math.}
\joudef{\jmp}    { J.\ Math.\ Phys.}
\joudef{\jpa}    { J.\ Phys.\ A}
\joudef{\mnras}  { Mon.\ Not.\ R.\ Ast.\ Soc.}
\joudef{\mpla}   { Mod.\ Phys.\ Lett.\ A} 
\joudef{\nature} { Nature}
\joudef{\nc}     { Nuovo Cim.}
\joudef{\npb}    { Nuc.\ Phys.\ B}
\joudef{\ph}     { Physica}
\joudef{\pla}    { Phys.\ Lett.\ A}
\joudef{\plb}    { Phys.\ Lett.\ B}
\joudef{\pr}     { Phys.\ Rev.}
\joudef{\prd}    { Phys.\ Rev.\ D}
\joudef{\prep}   { Phys.\ Rep.}
\joudef{\prl}    { Phys.\ Rev.\ Lett.}
\joudef{\prsla}  { Proc.\ Roy.\ Soc.\ Lond.\ A}
\joudef{\ptp}    { Prog.\ Theor.\ Phys.}
\joudef{\ptps}   { Prog.\ Theor.\ Phys.\ Suppl.}
\joudef\rmp      { Rev.\ Mod.\ Phys.}
\joudef\spj      { Sov.\ Phys.\ JETP}
\joudef\jetpl    { JETP Lett.}

\catcode`@=11

\newcommand\eqalign[1]{\null\,\vcenter{\openup\jot\m@th
  \ialign{\strut\hfil$\displaystyle{##}$&$\displaystyle{{}##}$\hfil
      \crcr#1\crcr}}\,}
\newcommand\meqalign[1]{\null\,\vcenter{\openup\jot\m@th
  \ialign{\strut\hfil$\displaystyle{##}$&&$\displaystyle{{}##}$\hfil
      \crcr#1\crcr}}\,}
\def\ps@reportnumber{%
    \let\@oddfoot\@empty\let\@evenfoot\@empty
    \def\@oddhead{\hfil\rightmark}}
	
\catcode`@=12   

\newdimen\arrayruleHwidth
\makeatletter
\newcommand\Hline{\noalign{\ifnum0=`}\fi\hrule \@height \arrayruleHwidth
  \futurelet \@tempa\@xhline}
\makeatother

\newcommand\thickbaselines{\baselineskip=20pt\lineskip=3pt\lineskiplimit=3pt}

\catcode`@=11

\renewcommand\matrix[1]{\null\,\vcenter{\thickbaselines\m@th
    \ialign{\hfil$##$\hfil&&\quad\hfil$##$\hfil\crcr
      \mathstrut\crcr\noalign{\kern-\baselineskip}
      #1\crcr\mathstrut\crcr\noalign{\kern-\baselineskip}}}\,} 
\catcode`@=12   

\renewcommand{\d}{{\rm d}} 

\newcommand{\arctanh}{\mathop{\rm arctanh}\nolimits}

\newcommand{\bo}{\boldsymbol}

\newcommand\bfv{{\bo v}}

\newcommand\undersim[1]{\mathop{\vtop{\ialign{##\crcr
     $\hfil\displaystyle{#1}\hfil$\crcr\noalign
     {\kern1pt\nointerlineskip}\hbox{$\hfil\sim\hfil$}\crcr
     \noalign{\kern1pt}}}}}

\newcommand{\acronym}[3]{\newcommand{#1}{#3 (#2)\relax\renewcommand{#1}{#2}}}



%% file: schw_simulation_letter.bbl
\begin{thebibliography}{1}

\bibitem{Novello_etal:2002}
{\em Artificial Black Holes}, edited by M. Novello, M. Visser, and G. Volovik
  (World Scientific Press, Singapore, 2002).

\bibitem{Visser:2003}
M. Visser, \ijmpd {\bf 12},  649  (2003).

\bibitem{Hau_etal:1999}
L. Vestergaard~Hau, S.~E. Harris, Z. Dutton, and C.~H. Behroozi, \nature {\bf
  397},  594  (1999).

\bibitem{Leonhardt&Piwnicki:2000}
U. Leonhardt and P. Piwnicki, \prl {\bf 84},  822  (2000).

\bibitem{Gordon:1923}
W. Gordon, \apb {\bf 72},  421  (1923).

\bibitem{Misner_etal:1973}
C.~W. Misner, K.~S. Thorne, and J.~A. Wheeler, {\em Gravitation} (Freeman, San
  Francisco, USA, 1973).

\bibitem{Robertson&Noonan:1968}
H.~P. Robertson and T.~W. Noonan, {\em Relativity and cosmology} (Saunders,
  London, 1968).

\bibitem{Martel&Poisson:2001}
K. Martel and E. Poisson, \ajp {\bf 69},  476  (2001),
  \eprintversion{gr-qc/0001069}.

\bibitem{Visser:1998}
M. Visser, \cqg {\bf 15},  1767  (1998).

\end{thebibliography}
